\documentclass{elsart}

\newcommand{\be}{\begin{equation}}
\newcommand{\ee}{\end{equation}}
\newcommand{\ben}{\begin{eqnarray}}
\newcommand{\een}{\end{eqnarray}}

\usepackage{amssymb}
\usepackage{graphicx}

\begin{document}
\begin{frontmatter}
\title{Tsallis' deformation parameter $q$ quantifies the classical-quantum
transition}
\author[fisica,CIC]{A.M. Kowalski},
\ead{kowalski@fisica.unlp.edu.ar}
\author[fisica,CONICET]{M. T. Martin},
\ead{mtmartin@fisica.unlp.edu.ar}
\author[fisica,CONICET]{A. Plastino\corauthref{cor}} and
\ead{plastino@fisica.unlp.edu.ar}
\author[ciop,ingenieria,fisica]{L. Zunino}
\ead{lucianoz@ciop.unlp.edu.ar}

\corauth[cor]{Corresponding author. Phone / Fax.: +54-11-4786 8114 }
\address[fisica]{Instituto de F\'isica (IFLP-CCT-Conicet), Fac. de Ciencias Exactas,\\
Universidad Nacional de La Plata,  C.C. 727, 1900 La Plata,
Argentina}
\address[CIC]{Comision de Investigaciones Cient\'ificas (CIC)}
\address[CONICET]{Argentina's National Research Council (CONICET)}
\address[ciop]{Centro de Investigaciones \'Opticas.\\
C.C. 124 Correo Central. 1900 La Plata, Argentina.}
\address[ingenieria]{Departamento de Ciencias B\'asicas, Facultad de Ingenier\'ia,\\ Universidad Nacional de La Plata (UNLP). 1900 La Plata, Argentina.}
\begin{abstract}

 We investigate the classical limit of  a type of
semiclassical evolution, the pertinent system representing the
interaction between matter  and a given field. On using as a
quantifier of the ensuing dynamics Tsallis q-entropy, we encounter
that it not only  appropriately describes the quantum-classical
transition, but that  the associated deformation-parameter q
itself characterizes the different regimes involved in the
process, detecting the most salient fine details of the
changeover.

PACS: 89.70.Cf. 03.65.Sq, 05.45.Mt

KEYWORDS: Tsallis entropy, Semiclassical theories, quantum chaos,
statistical complexity.

VERSION: 6.0
\end{abstract}
\end{frontmatter}
\section{Introduction}
\label{sec:Intro}
\subsection{Info-quantifiers}
Quantifiers based on information theory, like entropic forms and
statistical complexities (see as examples
\cite{Shannon48,Shiner99,LMC95,Lamberti04}) have proved to be quite
useful in the characterization of the dynamics associated to time
series, in the wake of the pioneering work of  Kolmogorov and Sinai,
who converted Shannon's information theory into a powerful tool for
the study of dynamical systems \cite{Kolmogorov58,Sinai59}.

Information theory measures and probability spaces $\Omega$ are
inextricably  linked. In the evaluation of above mentioned
quantifiers, the determination of the probability distribution $P$
associated to the dynamical system or time series under study is the
basic element. Many procedures have been proposed for the election
of $P \in \Omega$. We can mention techniques based on amplitude
statistics \cite{MT-Tesis}, symbolic dynamics \cite{Mischaikow99},
Fourier analysis \cite{Powell79}, and wavelet transform
\cite{Colo2001} (among others). The applicability of these
approaches depends on data-characteristics, i.e., stationarity,
length of the series, parameter-variations, levels of noise
contamination, etc. The distinct treatments ``capture'' the global
aspects of the dynamics, but they are not equivalent in their
ability to discern physical details. However, one should recognize
that we are here referring  to techniques defined in an ad-hoc
fashion, not derived directly from the dynamical properties of the
pertinent system itself.
\subsection{Deformed q-statistics}
It is  a well-known fact that physical systems that are
characterized by either long-range interactions, long-term memories,
or multi-fractality are best described  by a generalized statistical
mechanics' formalism \cite{Hanel} that was proposed 20 years ago,
being usually alluded to as deformed q-statistics. More precisely,
Tsallis~\cite{paper:tsallis1988} advanced in 1987 the idea of using
in a thermodynamics' scenario an entropic the Harvda-Chavrat form,
known today as Tsallis'  q-entropy, characterized by the entropic
index $q \in \mathcal{R}$ ($q \ne 1$):
\begin{equation}
S_q = \frac{1}{(q-1)} \sum _{i=1}^{N_s} \left[p_i - (p_i)^q\right],
\label{eq:t-entropy}
\end{equation}
where $p_i$ are the probabilities associated  with the associated
$N_s$ different system-configurations. The entropic index (or
deformation parameter) q describes the deviations of Tsallis entropy
from the standard Boltzmann-Gibbs-Shannon-one. Moreover, in the
limit $q \to 1$ Tsallis' entropy reduces to
 Shannon's \footnote{When $q \to 1$, $p_i^{q-1} =
e^{(q-1)\ln(p_i)} \sim 1+(q-1)\ln(p_i)$.}
\begin{equation}
S = - \sum _{i=1}^{N_s} p_i \ln (p_i). \label{Shannon-entropy}
\end{equation}
It is well-known that the orthodox entropy works best in dealing
with systems composed of either independent subsystems or
interacting via short-range forces whose subsystems can access all
the available phase space \cite{Hanel}. For systems exhibiting
long-range correlations, memory, or fractal properties,  Tsallis'
entropy becomes the most appropriate mathematical
form~\cite{paper:alemany1994,paper:tsallis1995,paper:capurro1998,paper:tsallis1998,paper:tong2002,paper:tsallis2003,paper:rosso2003b,paper:borland2005,paper:huang2005,paper:perez2007,paper:kalimeri2008}.

\subsection{Quantum-classical frontier}
The classical limit of quantum mechanics (CLQM) continues attracting
the attention of many theorists and is the source of much exciting
discussion (see, for instance, Refs.~\cite{paz,emerson} and
references therein). In particular, the investigation of ``quantum"
chaotic motion is considered important in this limit. Recent
literature provides us with many examples, although the adequate
definition of the underlying phenomena is understood in diverse
fashion according to the different authors (see Ref.~\cite{bifu} and
references therein).

Our motivation derives from the fact that it should be reasonable to
relay on q-statistics so as to gather insights into the

\be \label{suruta} {\rm
quantum\,\,-\,\,semiclassical\,\,-\,\,classical\,\, transition}. \ee
Why? Because we know that the classic to quantum route traverses
high complexity regions of the appropriate phase space where chaos
reigns, interrupted often by quasi-periodic windows
\cite{bifu,wavelet1,wavelet2}. In the semiclassical parcel of the
associated trajectory one encounters also strong correlation between
classical and quantum degrees of freedom \cite{wavelet1,wavelet2}.
The purpose of the present effort is precisely that of investigating
the possible q-statistics' contribution to this problem. Since in
this work the pertinent q-quantifiers are computed using ``wavelet
techniques'', we provide a brief wavelet-r\'esum\'e in the Appendix.

\section{A semi-classical model and the CLQM}
\label{Limite-clasico}

Quite a bit of quantum insight is to be gained from semiclassical
perspectives. Several methodologies are available (WKB,
Born-Oppenheimer approach, etc.) Here we consider two interacting
systems: a  classical and a quantal ones.
 This can be done whenever the quantum effects of one of the two systems
 are negligible in comparison to those of the other one.
 Examples can be readily found. We can just mention
  Bloch-equations \cite{Bloch}, two-level
 systems interacting with an electromagnetic field within a cavity,
   Jaynes-Cummings  semiclassical model \cite{Meystre,Bertsch,Bulgac,Milonni1,Milonni2,Kociuba}, collective nuclear motion
\cite{Ring}, etc.

More recently  \cite{Bonilla,Cooper2,Cooper3}, a special bipartite
model has been employed with reference to problems in such diverse
fields as chaos, wave-function collapse, measurement processes,
and cosmology \cite{Chung}. In order to tackle the problem posed
at the very end of the Introduction we shall consider a trivial
generalization of the semi-classical hamiltonian that represents
the zero-th mode contribution of a strong external field to the
production of charged meson pairs \cite{Cooper2,Cooper3}. It reads
\begin{equation}
\hat{H}~=~\frac{1}{2}\left(~\frac{\hat{p}^{2}}{m_{q}}~+~
                            \frac{{P_{A}}^{2}}{m_{cl}}~+~
                            m_{q}\omega ^{2}\hat{x}^{2}~\right) \ ,
\label{H}
\end{equation}
where {\it i)\/}   $\hat{x}$ and $\hat{p}$ are quantum operators,
{\it ii)\/}  $A$ and $P_{A}$ classical canonical conjugate
variables and {\it iii)\/} $\omega ^{2}={\omega
_{q}}^{2}+e^{2}A^{2}$ is an interaction term that introduces
nonlinearity, $\omega _{q}$ being a frequency. The quantities
$m_{q}$ and $m_{cl}$ are masses, corresponding to the quantum and
classical systems, respectively. As shown in  Ref.~\cite{present},
in dealing with   (\ref{H}) one faces an autonomous system of
nonlinear coupled equations
\begin{equation}
\begin{tabular}{lll}
$\frac{d\langle \hat{x}^{2}\rangle }{dt}=\frac{\langle \hat{L}\rangle }{m_{q}} \ , \qquad $
& $\frac{d\langle \hat{p}^{2}\rangle }{dt}=-m_{q}\ \omega ^{2}\langle \hat{L}\rangle \ , \qquad $
& $\frac{d\langle \hat{L}\rangle }{dt}    =2( \frac{\langle \hat{p}^{2}\rangle }{m_{q}}-m_{q} \
\omega ^{2}\langle \hat{x}^{2}\rangle)$, \\
$\frac{dA}{dt}= \frac{P_{A}}{m_{cl}}                          \ , \qquad $ &
$\frac{dP_{A}}{dt}=-e^{2}m_{q}\,A\langle \hat{x}^{2}\rangle   \ , \qquad $ &
$\hat{L}=\hat{x} \hat{p}+ \hat{p} \hat{x}$ \ . \\
\end{tabular}
\label{eqq}
\end{equation}
The system of Eqs. (\ref{eqq}) follows immediately from
Ehrenfest's relations \cite{present}. To study the classical limit
we need to also consider the classical counterpart of  the
Hamiltonian (\ref{H})
\begin{equation}
H~=~\frac{1}{2}\left[\frac{p^{2}}{m_{q}}~+~
                     \frac{{P_{A}}^{2}}{m_{cl}}~+~
                     m_{q}(\omega_{q}^{2}+e^{2}A^{2})x^{2}
              \right] \ ,
\label{Hcl}
\end{equation}
where all the variables are classical.
Recourse to Hamilton's equations allows one to find the classical version of Eqs.
(\ref{eqq}) (see  Ref.~\cite{present} for details).
The classical limit is obtained by letting the ``relative energy''
\begin{equation}
E_{r}~=~ \frac{|E|}{I^{1/2} \omega _{q}} \rightarrow \infty,
\label{limita}
\end{equation} where $E$ is the total energy of the system and $I$ an invariant
of the  motion described by the system (\ref{eqq}), related to the
Uncertainty Principle
\begin{equation}
I ~=~ \langle \hat{x}^{2}\rangle \langle \hat{p}^{2}\rangle -\frac{\langle \hat{L}\rangle ^{2}}{4}.
\end{equation}
A classical computation of $I$ yields $I=x^{2}p^{2}-L^{2}/4\equiv
0$. A measure of the convergence between classical and quantum
results in the limit of Eq. (\ref{limita}) is given by the norm
${\mathcal N}$ of the vector $\Delta u=u-{u_{cl}}$ \cite{present}
\begin{equation}
{\mathcal N}_{\Delta u}=| u-{u_{cl}} | \ ,
\label{norma}
\end{equation} where the three components vector $u=(\langle \hat{x}^{2}\rangle ,\langle \hat{p}^{2}\rangle ,\langle \hat{L}\rangle )$
is the ``quantum'' part of the solution of the system Eqs.
(\ref{eqq}) and $u_{cl}=(x^{2},p^{2},L)$ its classical partner.

A detailed study of this model, was performed in
Refs.~\cite{present,pla}. We summarize here the main results of
these references that are  pertinent for our discussion. In
plotting diverse dynamical quantities versus $E_r$ (as it grows
from unity to $\infty$), one finds {\it an abrupt change in the
system's dynamics for special values of $E_{r}$, to be denoted by
${E_{r}}^{cl}$}. From this value onwards, the pertinent dynamics
starts converging to the classical one. It is thus possible to
assert that ${E_{r}}^{cl}$ provides us with an {\it indicator} of
the presence of a quantum-classical ``border". The zone
\begin{equation}
E_{r}~<~{E_{r}}^{cl} ,
\end{equation}
corresponds to the semi-quantal regime investigated in
Ref.~\cite{pla}. This regime, in turn, is characterized by {\it two}
different sub-zones  \cite{present}. {\it i)\/} One of them is an
almost purely quantal one, in which the microscopic quantal
oscillator is just slightly perturbed by the classical one, and {\it
ii)\/} the other section exhibits a transitional nature
(semi-quantal). The border between these two sub-zones can be well
characterized by a ``signal'' value ${E_{r}}^{{\mathcal P}}$.
 A significant feature of this point resides in the fact that, for  $E_{r}\geq {E_{r}}^{{\mathcal P}}$, {\it chaos is always found}. The relative number of
  chaotic orbits (with respect to the total number of
orbits) grows with $E_r$ and tends to unity for $E_{r}\rightarrow
\infty$ \cite{present,pla}.

Thus, as $E_r$ grows from $E_{r}=1$ (the ``pure quantum instance")
to $E_{r}\rightarrow \infty $ (the classical situation), a
significant series of {\it morphology-changes} is detected,
specially in the transition-zone (${E_{r}}^{{\mathcal P}}\leq E_r
\leq {E_{r}}^{cl}$). The concomitant orbits exhibit features that
are not easily describable in terms of Eq. (\ref{norma}), which is a
{\it global} measure of convergence in amplitude (of the signal).
What one needs instead is a statistical type of characterization, as
that described in  Refs. \cite{wavelet1,wavelet2,BP}. In the present
work we will present a different, novel perspective of the
quantum-classical transition problem and a characterization of it
that we believe to be more insightful and of a stronger convincing
nature than those of \cite{wavelet1,wavelet2,BP}.

\section{Present results}

By recourse to the so-called  normalized Tsallis wavelet entropy (NTWE)
$\mathcal{H}_{S_q}$ given by (see Appendix)
\begin{equation}
\mathcal{H}_{S_q}[P]=S_{q}[P]/S_{q,{max}}
        = \frac{1}{1-N_J^{1-q}} \sum _{j=-1}^{-N_J} \left(p_{j} - p_j^q\right),
\label{Tsa1}
\end{equation} we will be able to characterize the details that pave
the road towards  the classical limit.

In obtaining our numerical results we choose $m_{q}= m_{cl}=
\omega_{q}= e = 1$ for the system's parameters. As for the initial
conditions for solving the system (\ref{eqq}) we take $E= 0.6$,
i.e., we fix $E$ and then vary $I$ so to obtain our different
$E_r$-values. Additionally, we have $\langle L \rangle(0)= L(0)=0$
and $A(0)=0$ (both in the quantum and the classical instances).
$\langle x^2 \rangle(0)$ takes values in the interval $x^2 (0) <
\langle x^2 \rangle(0)\leq 0.502$, with $x^2(0)= 0.012$.

We define eight ($N_J=8$) resolution levels $j = -1, -2, \cdots ,-N_J$ for an appropriate wavelet analysis
within the multiresolution scheme. The $p_j$
yield, at different scales, the energy probability distribution and so NTWE constitutes a
suitable tool for detecting and characterizing specific phenomena.

We find as first result that the range of $q$ values that allows for
a correct description, is $0 < q <5$ (see Figs. 1, 2 and 3, where we
depicts the normalized Tsallis q entropy $\mathcal{H}_{S_q}$ vs.
$E_r$ for different q-values). In this range $\mathcal{H}_{S_q}$
distinguishes the three sections of our process, i.e., quantal,
transitional, and classic, as delimited by, respectively,
${E_{r}}^{{\mathcal P}}=3.3282$ and ${E_{r}}^{cl}=21,55264$. Notice
the abrupt change of in the slope of the curve taking place at
${E_{r}}^{{\mathcal P}}$, where a local minimum is detected for
$q>0.1$ (Fig 1a). The transition zone is clearly demarcated between
that point and ${E_{r}}^{cl}$. From here onwards $\mathcal{H}_{S_q}$
adopts a ``horizontal'' behavior as it tends to its classical value
at the same time that the solutions of (\ref{eqq}) begin to converge
towards the classical ones. If $q\geq5$  two of our  zones:
transitional and classical, lose their identity (see below).

In general, the most noticeable $\mathcal{H}_{S_q}-$ changes take
place in the quantal zone,  specially for $q<1$ (Figs. 1a-b) and in
the frontier between the transitional and classical zones, now for
$q>3$ (Fig. 3).

The $\mathcal{H}_{S_q}-$slope changes in the quantal sector as $q$
varies, being negative for $q>07$ (Figs. 1b, 2, and 3) and
positive for $0<q<0.7$, if $E_r \leq 2.6$ (Figs. 1a). The slope is
null in this last $E_r-$interval, for $q\simeq0.7$ (Fig. 1b).
Thus, for $q \le \sim 0.4$, $\mathcal{H}_{S_q}-$values are, in the
quantal zone, smaller (or equal) to those pertaining to the
transitional one.

One concludes then that in the interval $0<q\leq0.4$,
$\mathcal{H}_{S_q}$ {\it is able to represent in better fashion
the quantal zone' features than Shannon's entropy} (see
\cite{BP}), as the information measure should be smaller in this
zone than in the transitional one.
 In view of the meaning assigned to ${E_{r}}^{{\mathcal P}}$, it is
then important to have a $\mathcal{H}_{S_q}-$minimum there. These
considerations allow us to conclude that the optimal $q-$range is
$0.1<q\leq0.4$.

For $0.7\leq q<1$, and as $q\rightarrow 1$, the q-entropy behavior
resembles more and more the one of  Shannon's measure. No great
changes ensue for $1<q<3$, while for  $3\leq q<5$,  the sharp
demarcation of our three sections starts deteriorating. Notice in
Fig. 3 that as $q$ grows the $\mathcal{H}_{S_q}-$curve starts
acquiring a ``horizontal'' nature in the transition region close to
the classical one. A ``merging" between the two sectors takes place
for $q\geq5$.

The $q-$influence on these processes is seen in  Figs. 4, that plot
$\mathcal{H}_{S_q}$ vs. $q$ for different values of our all
important quantity $E_r$.  The corresponding Shannon entropy value
(horizontal line) is included in all graphs for comparison's
purpose.  Typical kinds of morphology are exhibited by these figures
\cite{Luciano}. Figs. 4a)-4b) correspond to the quantum sector,
while Figs. 4c), 4d), 4e), and 4f) refer to  the transitional one,
and, finally, Figs. 4g)-4h) allude to the classical region. We
verify that $\mathcal{H}_{S_q}$
 possesses only one minimum as a function of $q$. Consequently,  $\mathcal{H}_{S_q}$ intersects  Shannon's curve at
 two points, i.e.,  i) $q=1$ and ii) another $q-$value, $q^*$, that   depends on
 $E_{r}$. Note that  $q^*<1$ always, save for the $E_r={E_{r}}^{{\mathcal P}}$ instance (in this case $q^*\simeq 2.55$).
  Figure 4c) depicts precisely this peculiar situation. The protagonist of Figs. 4d)-4e) is
    $\mathcal{H}_{S_q}$ ``tangency" to the Shannon
   entropy. Figure 4e) corresponds to $E_r=6,81554$, a point that divides into two
   sections the transitional
   region, one in which the quantum-classical mixture characterizes  a phase-space
with more non-chaotic than chaotic curves and other, in which this
aspect is reversed  \cite{pla}. That latter feature is also typical
of the  classical section. Comparison between figures 4f)
(transitional sub-zone) and 4g)-4h) (classical area)
exhibits the coherence of the above line of discourse.

 Thus,
$\mathcal{H}_{S_q}$ as a function of  $q$ is perfectly able, by
itself, i) of ``detecting" important dynamical features like the
``Signal Point'' ${E_{r}}^{{\mathcal P}}$ and ii) of
distinguishing between the two transitional sub-regions and
registering the similarities between the second of these two and
the classical one. Remarkably enough, if we disregard now the
transitional zone and just
 compare the quantal with the classical ones,
some similarities and differences are also evidenced. The facts
described in this paragraph demand for the finding of a quantity
that might provide one with a way of giving quantitative clothing to
the qualitative features just considered.

 We propose here  that, in this respect, a useful quantity is the
$\mathcal{H}_{S_q}-$curvature (HC)
\begin{equation}\label{curvature}
    \kappa(q, E_r)=\frac{\mid\frac{\partial^2 \mathcal{H}_{S_q}}{\partial q^2}\mid}{\left(1+(\frac{\partial \mathcal{H}_{S_q}}{\partial
    q})^2\right)^\frac{3}{2}}.
\end{equation} Fig. 5, depicts $\kappa(q_M, E_r)$ vs.
$E_r$, evaluated for that $q-$value, $q_M$, minimizing
$\mathcal{H}_{S_q}$ (yielding a  value $(\mathcal{H}_{S_q})_M$).
We appreciate that HC's values   differ in our three zones:
quantal, transitional , and  classical. The three of them are more
sharply delineated using the HC $\kappa$ than by recourse to
$\mathcal{H}_{S_q}$, particularly with regards to the quantum
region.

In figures 6 we display the quantities $q_M$ and
$(\mathcal{H}_{S_q})_M$ versus $E_r$. The
$(\mathcal{H}_{S_q})_M-$graph (in which $q$ is different at every
$E_r$) of Fig. 6a) astonishingly resembles those of
$\mathcal{H}_{S_q}$ vs. $E_r$. In turn, $q_M$ (Fig. 6b) behaves
clearly as a ``demarcator" that performs an exceptionally good job
at exhibiting the convergence towards classicality.

\textbf{One is thus led to the  conclusion that it is the parameter
$q$ itself
 the one detecting the quantum-classical transition, via different $q-$dependent quantities}.

\section{Conclusions}

The focus of attention in this communication has been the
classical-quantal frontier, as looked through the glass of a
wavelet-band analysis and by recourse to the dynamics governed by
a semi-classical hamiltonian that represents the zero-th mode
contribution of an strong external field to the production of
charged meson pairs. This study is in turn encompassed within
Tsallis' statistics.

The highlights of the road towards classicality are described by
recourse to the relative energy $E_r$ given by (\ref{limita}). As
$E_r$ grows from $E_{r}=1$ (the ``pure quantum instance") to
$E_{r}\rightarrow \infty $ (the classical situation), a
significant series of {\it morphology-changes} is detected for the
solutions of the system of nonlinear coupled equations
(\ref{eqq}). The concomitant process takes place in three stages:
 quantal, transitional, and classic, delimited, respectively, by
special values of  $E_r$, namely, ${E_{r}}^{{\mathcal P}}$ and
${E_{r}}^{cl}$.

We were able to ascertain that the normalized Tsallis wavelet
entropy $\mathcal{H}_{S_q}$,  in the range $0 < q <5$, correctly
describes the dynamical $E_r-$evolution, unmistakably identifying
the above mentioned three stages. As a {\it second} result we have
ascertained that within the subrange $0.1 < q \leq 0.4$,
$\mathcal{H}_{S_q}$, not only {\it identifies} the three different
$E_r-$regions but also properly portrays the quantum sector,
something that Shannon's measure is unable to do. We have thus
encountered that the wavelet-constructed $\mathcal{H}_{S_q}$, in the
range $0.1 < q \leq0.4$, is the most appropriate entropy, and not
the orthodox, $q=1$ of Shannon's.

Thirdly, we find that  $\mathcal{H}_{S_q}$, as a function of $q$
 is a good ``detector" of transitional features (see Figs. 4): a) identifies ${E_{r}}^{{\mathcal P}}$, starting point of the
 transitional sector and where chaotic behavior  begins to emerge and b) distinguishes between the two subsections
 into which the transitional region divides itself:
one in which the quantum-classical mixture characterizes a
phase-space with more non-chaotic than chaotic curves and other,
in which this aspect is reversed.

Finally,  we have discovered other transition-detectors in
addition to the normalized Tsallis-entropy, specially its
curvature when we plot it for that particular $q-$value $q_M$ for
which $\mathcal{H}_{S_q}$ has a minimum (Fig. 5).  $q_M$ itself
(Fig. 6b) turns out to a good transition-indicator. These last
results affirm  that the Tsallis parameter $q$ by itself can be
regarded as the ``looking glass'' through which one can observe
the quantum-classical transition.

\section{Acknowledgments}

AMK are supported by CIC of Argentina.

\appendix
\section{Normalized Tsallis wavelet entropy}
\label{sec:Wavelet-Transform} Wavelet analysis is a suitable tool
for detecting and characterizing specific phenomena in time and
frequency planes. The {\it wavelet} is a smooth and quickly
vanishing oscillating function with good localization in both
frequency and time.

A {\it wavelet family} $\psi _{a,b}( t )~=~|a|^{-1/2}\psi \left(
{{t-b}\over {a}} \right)$ is the set of elementary functions
generated by dilations and translations of a unique admissible {\it
mother wavelet} $\psi (t)$. $a, b \in {\mathcal R}$, $a \not = 0$
are the scale and translation parameters respectively, and $t$ is
the time. One have a unique analytic pattern and its replications at
different scales and with variable time localization.

For special election of the mother wavelet function $\psi(t)$ and
for the discrete set of parameters, $a_{j} = 2^{-j}$ and $b_{j,k} =
2^{-j} k$,  with $j, k \in {\mathcal Z}$ (the set of integers) the
family
\begin{equation}
\label{eq:wav3} \psi_{j,k}( t )~=~2^{j/2}~\psi(~2^j~t~-~k~) \quad
\quad \quad j,~k~\in~{\mathcal Z}  \  ,
\end{equation}
constitutes an orthonormal basis of the Hilbert space $L^2({\mathcal
R})$ consisting of finite-energy signals.

The correlated {\it decimated discrete wavelet transform\/}
provides a non-redundant representation of the signal $X$, and the
values $\left<~{\mathcal X},~\psi_{a,b}~\right>$ constitute the
coefficients in a wavelet series. These wavelet coefficients
provide relevant information in a simple way and a direct
estimation of local energies at the different scales. Moreover,
the information can be organized in a hierarchical scheme of
nested subspaces called multiresolution analysis in $L^2({\mathcal
R})$. In the present work, we employ orthogonal cubic spline
functions as mother wavelets. Among several alternatives, cubic
spline functions are symmetric and combine in a suitable
proportion smoothness with numerical advantages.

In what follows, the signal is assumed to be given by the sampled
values ${\mathcal X} = \{x(n), n = 1,\cdots, N \}$. If the
decomposition is carried out over all resolutions levels the wavelet
expansion will read ($N_J = \log_2(N)$)
\begin{equation}
\label{eq:wav4} {\mathcal S}( t ) ~=~ \sum_{j=
-N_J}^{-1}~\sum_k~C_j(k)~\psi_{j,k}(t)
                  ~=~ \sum_{j= -N_J}^{-1}~r_j(t)   \  ,
\end{equation}
where the wavelet coefficients $C_j(k)$ can be interpreted as the
local residual errors between successive signal approximations at
scales $j$ and $j+1$, and $r_{j}(t)$ is the {\it residual signal} at
scale $j$. It contains the information of the signal ${\mathcal
S}(t)$ corresponding to the frequencies $2^{j-1} \omega_s \leq |
\omega | \leq 2^j \omega_s$.

Since the family $\{ \psi_{j,k}(t) \}$ is an {\it orthonormal} basis
for $L^2({\mathcal R})$, the concept of energy is linked with the
usual notions derived from  Fourier's theory. The wavelet
coefficients are given by $C_j(k) = \left< {\mathcal S}, \psi_{j,k}
\right>$ and the energy, at each resolution level $j= -1, \cdots,
-N_J$, will be the energy of the detail signal
\begin{equation}
E_j~=~\|r_j\|^2~=~\sum_k~|C_j(k)|^2. \label{eq:wav6}
\end{equation}
The total energy can be obtained in the fashion
\begin{equation}
\label{eq:wav7} E_{tot}~=~\| {\mathcal S}
\|^2~=~\sum_{j<0}~\sum_k~|C_j(k)|^2
       ~=~\sum_{j<0}~E_j  \  .
\end{equation}
Finally, we define the normalized $p_j$-values, which represent the
{\it relative wavelet energy\/}
\begin{equation}
\label{eq:wav8} p_j~=~E_j~/~E_{tot}
\end{equation}
for the resolution levels $j = -1, -2, \cdots ,-N_J$. The $p_j$
yield, at different scales, the probability distribution for the
energy. Clearly, $\sum_{j} p_j = 1$ and the distribution $\{ p_j \}$
can be considered as a time-scale density that constitutes a
suitable tool for detecting and characterizing specific phenomena in
both the time and the frequency planes.

The normalized Tsallis wavelet entropy (NTWE) is just the normalized
Tsallis entropy associated to the probability distribution $P$,
\begin{equation}
\mathcal{H}_{S_q}[P]=S_{q}[P]/S_{q,{max}}
        = \frac{1}{1-N_J^{1-q}} \sum _{j=-1}^{-N_J} \left(p_{j} - p_j^q\right),
\label{Tsa}
\end{equation}
where $S_{q,{max}}= (1-N_J^{1-q})/(q-1)$ is attained for the
equiprobable distribution $P_e=\{1/N_J,\dots,1/N_J\}$.

The NTWE appears as a measure of the degree of order/disorder of the
time series. It provides useful information about the underlying
dynamical process associated with the series. Indeed, a very ordered
process can be represented by a periodic mono-frequency signal
(signal with a narrow band spectrum). A wavelet representation of
such a signal will be resolved at one unique wavelet resolution
level, i.e., all relative wavelet energies will be (almost) zero
except at the wavelet resolution level which includes the
representative series frequency. For this special level the relative
wavelet energy will (in our chosen energy units) almost equal unity.
As a consequence, the NTWE  will acquire a very small, vanishing
value. A signal generated by a totally random process or chaotic one
can be taken as representative of a very disordered behavior. This
kind of signal will have a wavelet representation with significant
contributions coming from all frequency bands. Moreover, one could
expect that all contributions will be of the same order.
Consequently, the relative wavelet energy will be almost equal at
all resolutions levels, and the NTWE  will acquire its maximum
possible value.

\begin{figure} [tbp]
\begin{center}
{\includegraphics{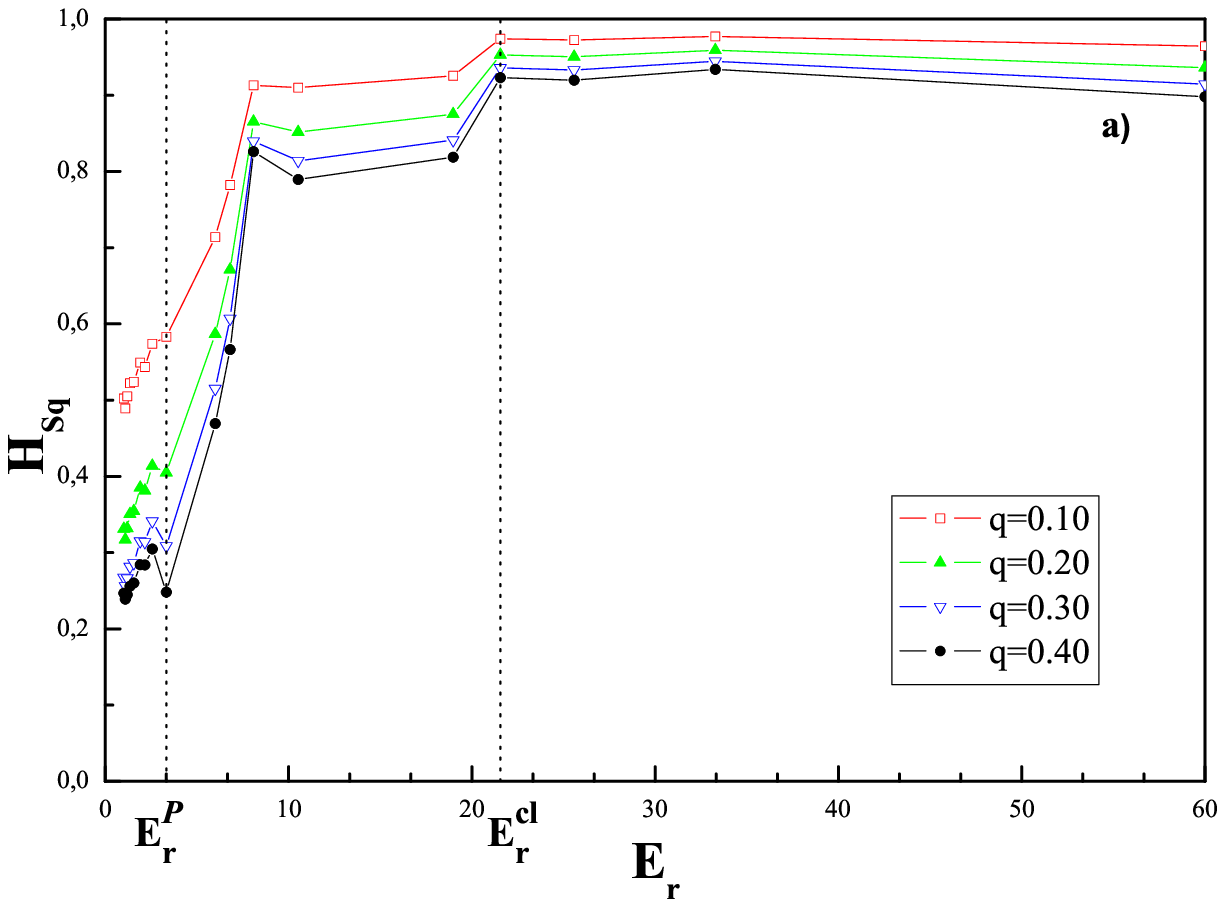}}
{\includegraphics{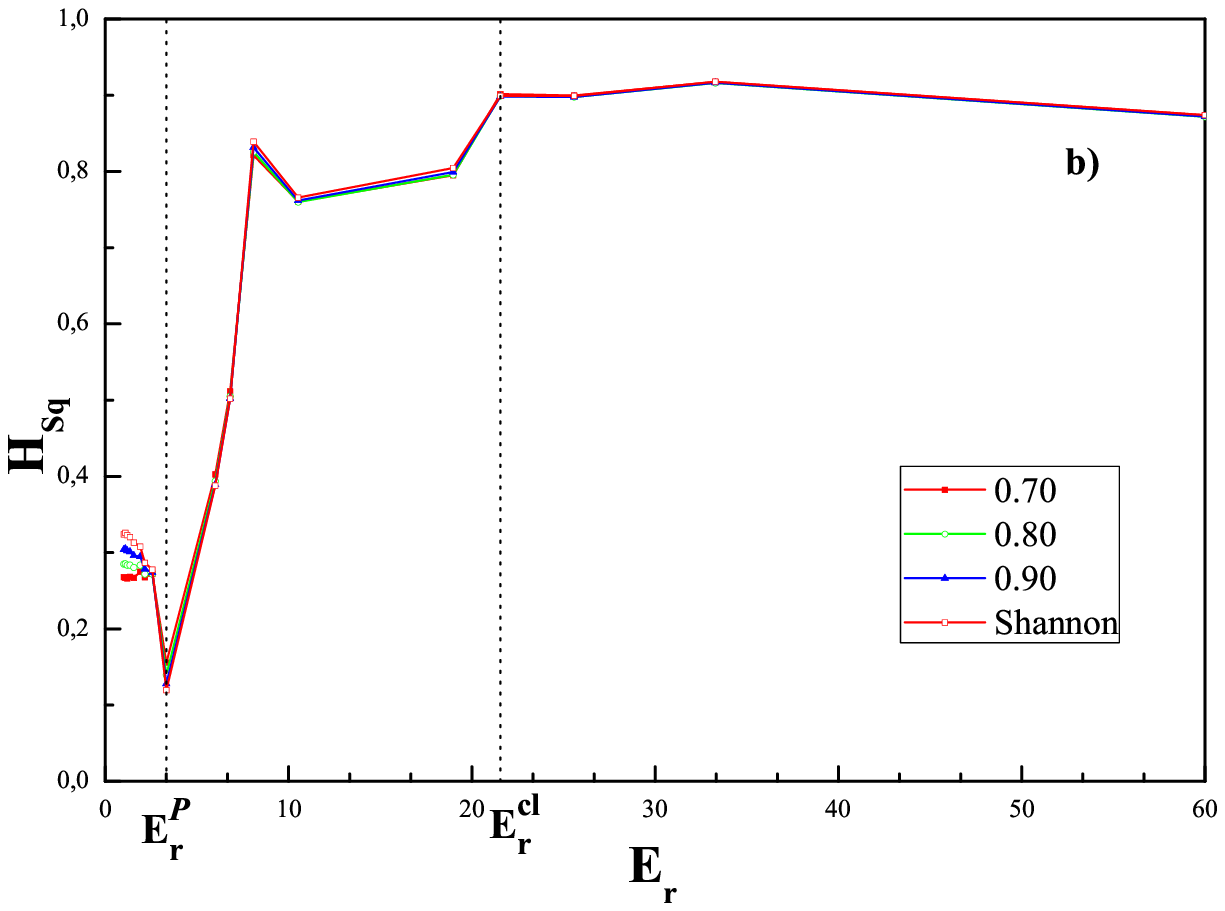}}
\caption{Normalized Tsallis entropy $\mathcal{H}_{S_q}$ vs. $E_r$
for  $q < 0.7$ (Fig. 1a) and $0.7 \leq q <1$  (Fig. 1b). Shannon's
entropies are also displayed. Three zones are to be differentiated.
They are delimited by  special $E_r-$values, namely,
${E_{r}}^{{\mathcal P}}=3.3282$ and ${E_{r}}^{cl}=21,55264$. Note
that the quantum sector is that for which $1\leq E_r
<{E_{r}}^{{\mathcal P}}$ and, there, $\mathcal{H}_{S_q}$'s slope
changes with $q$, being negative for $q>0.7$ (Figs. 1b) and positive
for $0<q<0.7$, if $E_r \leq 2.6$ (Figs. 1a). For $0.1<q\leq0.4$,
$\mathcal{H}_{S_q}-$values are in the quantal zone smaller than (or
equal to) in the transitional one.}
\end{center}
\end{figure}

\begin{figure} [tbp]
\begin{center}
{\includegraphics{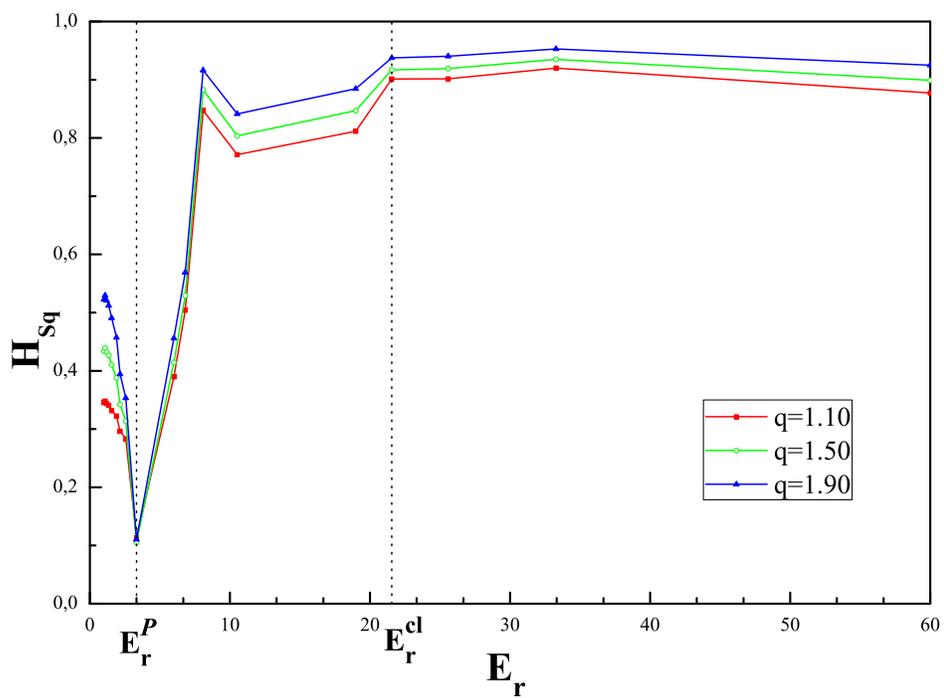}}
\caption{Normalized Tsallis entropy $\mathcal{H}_{S_q}$ vs $E_r$ for
$1<q \leq 2$. The three zones of Fig. 1 are also seen here.}
\end{center}
\end{figure}

\begin{figure} [tbp]
\begin{center}
{\includegraphics{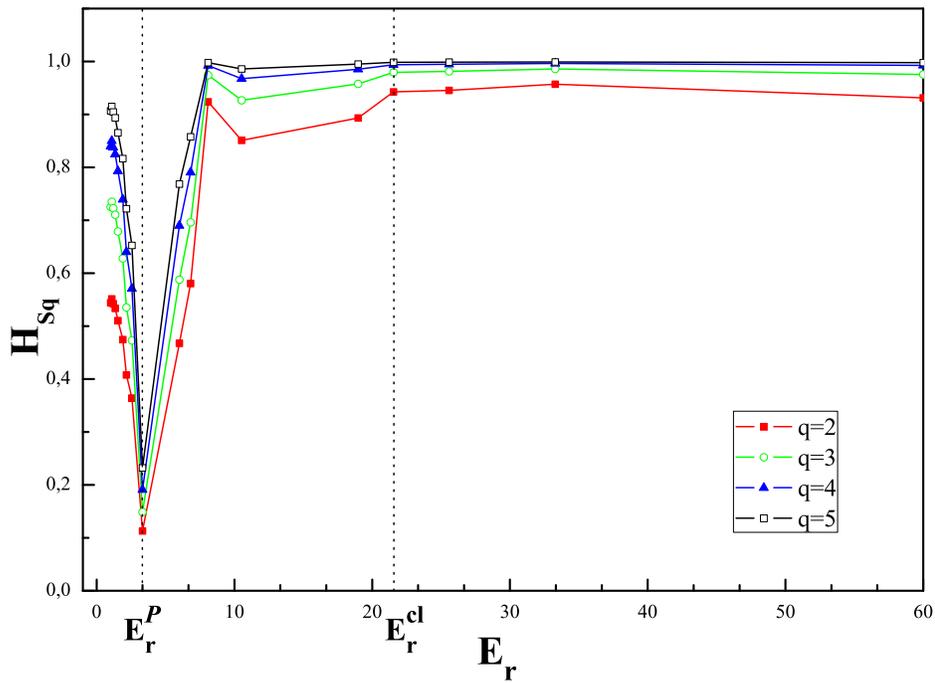}}
\caption{Normalized Tsallis entropy $\mathcal{H}_{S_q}$ vs. $E_r$
for  $2\leq q \leq 5$. In the range  $1<q<3$ the morphology is that
of Fig. 1. For $3\leq q<5$, the $\mathcal{H}_{S_q}-$curve starts
acquiring a ``horizontal" nature in the transition region close to
the classical one. For $q>5$, a ``merging'' between the two sectors
takes place.}
\end{center}
\end{figure}

\begin{figure} [tbp]
\begin{center}
{\includegraphics{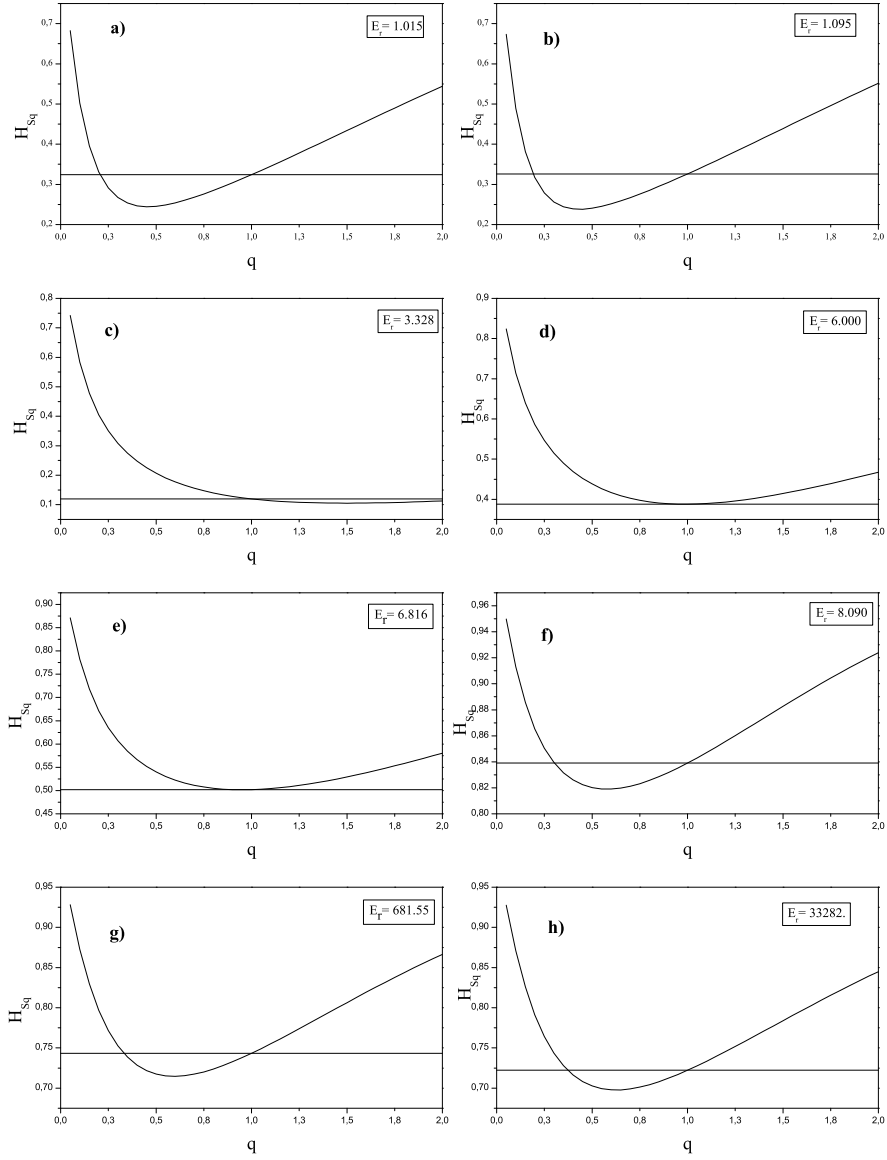}} \caption{Normalized Tsallis entropy
$\mathcal{H}_{S_q}$ vs. $q$ for different $E_r$-values.  1) Quantal
(Figs. 4a - 4b), transitional (Figs. 4c, 4d, 4e and 4f) and classic
(4g - 4h). The corresponding Shannon entropy value (horizontal line)
is included in all graphs for comparison's purpose. Note that
$\mathcal{H}_{S_q}$ intersects
 the Shannon entropy-curve for  $q>1$ in figure 4c) and becomes tangent to it
 in figures 4d) - 4e).}
\end{center}
\end{figure}

\begin{figure} [tbp]
\begin{center}
{\includegraphics{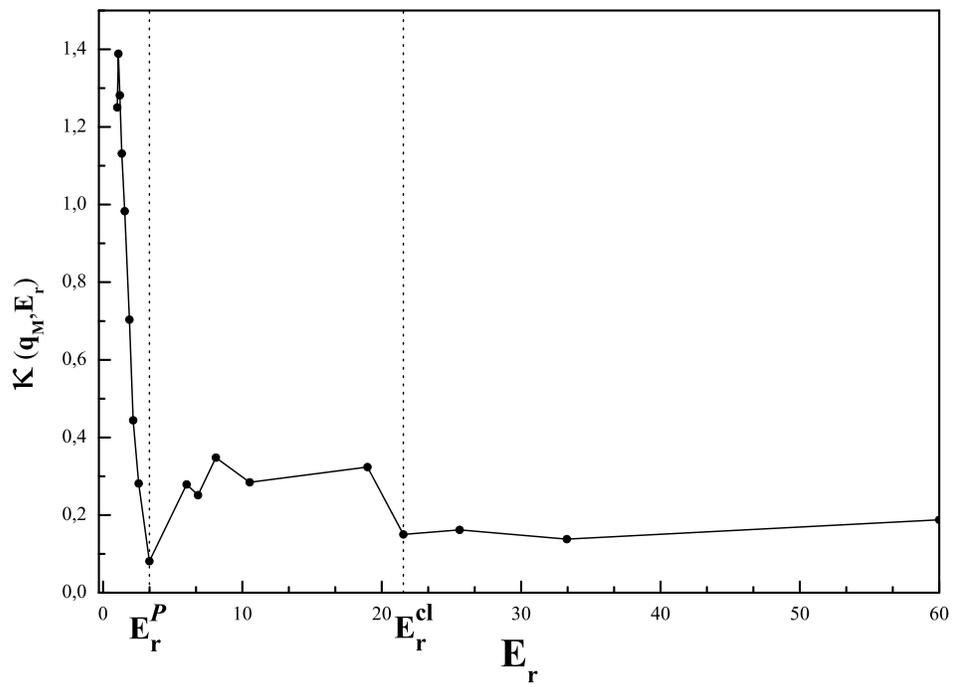}}
\caption{We depict the curvature $\kappa(q_M, E_r)$ vs. $E_r$. At
 $q=q_M$  $\mathcal{H}_{S_q}$ adopts its minimum value. The three regions of Fig. 1 are clearly delineated.
 }
\end{center}
\end{figure}

\begin{figure} [tbp]
\begin{center}
{\includegraphics{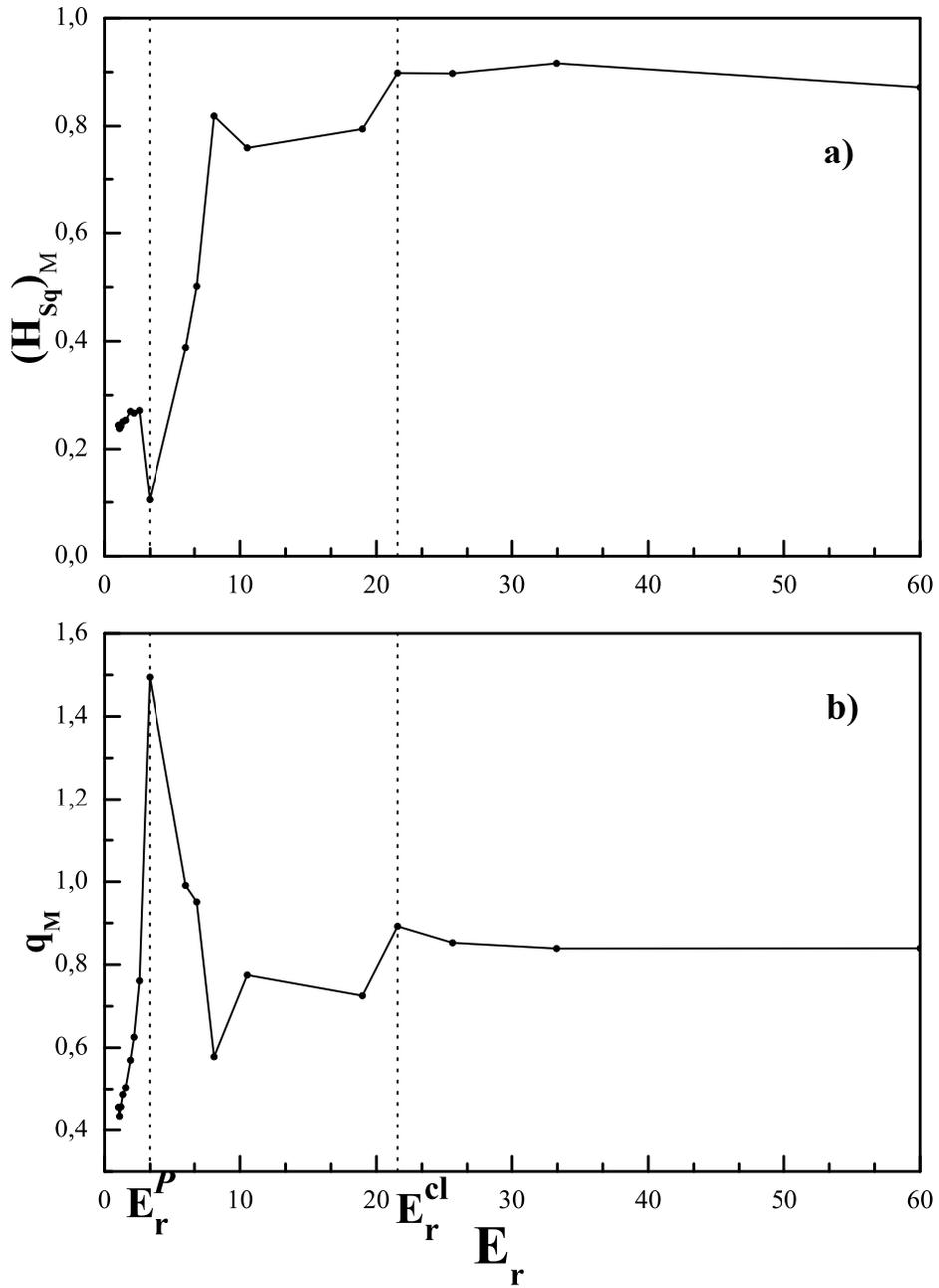}}
\caption{6a) Graph of $(\mathcal{H}_{S_q})_M$, the minimum
$\mathcal{H}_{S_q}-$value ,  versus $E_r$.  6b)  $q_M$, the particular
$q-$value at which that minimum is attained versus $E_r$. The two
plots display the quantum-classical transition.}
\end{center}
\end{figure}

\end{document}